\documentclass[runningheads]{llncs}

\usepackage{cite}
\usepackage[pdftex]{graphicx}
\usepackage[cmex10]{amsmath}
\usepackage{algorithmic}
\usepackage[ruled]{algorithm2e}
\usepackage{flushend}
\usepackage{listings}
\usepackage{booktabs}
\usepackage[hyphens]{url} 
\usepackage{pbox}

\usepackage{pifont}                        
\newcommand{\xmark}{\leavevmode{\color{red}\ding{55}}}
\newcommand{\cmark}{\leavevmode{\color{green!50!black}\ding{51}}}
\usepackage{floatrow}
\DeclareFloatFont{tiny}{\scriptsize}
\usepackage[utf8]{inputenc}
\usepackage{tikz}
\usepackage{multirow}
\usepackage{pgfplots}
\usetikzlibrary{arrows}
\usetikzlibrary{positioning}
\usetikzlibrary{decorations.pathreplacing}
\usetikzlibrary{pgfplots.groupplots}
\usetikzlibrary{external}
\usepackage{letltxmacro}
\pgfplotsset{compat=newest}
\usepackage{ifthenx}
\usepackage{paralist}
\usepackage{xspace}

\usepackage{csvenhanced}
\usepackage{siunitx,booktabs}

\usepackage{algorithmic}
\usepackage[ruled]{algorithm2e}

\usepackage{color}
\usepackage{paralist}
\lstdefinestyle{mystyle}{
    commentstyle=\color{green!40!black},
    basicstyle=\scriptsize\ttfamily,
    breakatwhitespace=false,
    breaklines=true,
    captionpos=b,
    keepspaces=true,
    numbers=left,
    numbersep=5pt,
    showspaces=false,
    showstringspaces=false,
    showtabs=false,
    tabsize=2,
    xleftmargin=15pt,
    numberbychapter=false
}
\newcommand{\etal}{et~al.\ }
\newcommand{\ie}{\textit{i.e.},\ }
\newcommand{\eg}{e.g.,\ }
\newcommand{\flushonly}{Flush+Flush\xspace}
\newcommand{\FlushOnly}{\emph{Flush+Flush}\xspace}
\newcommand{\clflush}{\texttt{clflush}\xspace}
\newcommand{\FlushReload}{\emph{Flush+Reload}\xspace}
\newcommand{\PrimeProbe}{\emph{Prime+Probe}\xspace}
\newcommand{\mceil}[1]{\left\lceil #1 \right\rceil}
\usepackage{siunitx,array,booktabs}

\begin{document}

\title{{\flushonly}:\\A Fast and Stealthy Cache Attack}

\author{Daniel Gruss \and Clémentine Maurice$^{\dagger}$ \and Klaus Wagner \and Stefan Mangard}
\institute{Graz University of Technology, Austria}

\maketitle

\begin{abstract}
Research on cache attacks has shown that CPU caches leak significant information.
Proposed detection mechanisms assume that all cache attacks cause more cache hits and cache misses than benign applications
and use hardware performance counters for detection.

In this article, we show that this assumption does not hold by developing a novel attack technique: the \FlushOnly attack.
The \FlushOnly attack only relies on the execution time of the flush instruction, which depends on whether data is cached or not. 
\FlushOnly does not make any memory accesses, contrary to any other cache attack. Thus, it causes no cache misses at all and the
number of cache hits is reduced to a minimum due to the constant cache flushes.
Therefore, \FlushOnly attacks are stealthy, \ie the spy process cannot be detected based on cache hits and misses, or state-of-the-art detection mechanisms.
The \FlushOnly attack runs in a higher frequency and thus is faster than any existing cache attack. With 496\,KB/s in a cross-core covert channel it is $6.7$ times faster than any previously published cache covert channel.
\end{abstract}

\section{Introduction}

\bgroup
\let\thefootnote\relax\footnotetext{This paper has been accepted at DIMVA 2016 (\url{dimva2016.mondragon.edu/en}). The final publication is available at link.springer.com (\url{http://link.springer.com/}).}
\let\thefootnote\relax\footnotetext{$^\dagger$ Part of the work was done while author was affiliated to Technicolor and Eurecom.}
\egroup

The CPU cache is a microarchitectural element that reduces the memory access time of recently-used data. It is shared across cores in modern processors, and is thus a piece of hardware that has been extensively studied in terms of information leakage. Cache attacks include covert and cryptographic side channels, but caches have also been exploited in other types of attacks, such as bypassing kernel ASLR~\cite{Hund2013}, detecting cryptographic libraries~\cite{Irazoqui2015Neighbor}, or keystroke logging~\cite{Gruss2015Template}.
Hardware performance counters have been proposed recently as an OS-level detection mechanism for cache attacks and Rowhammer~\cite{Chiappetta2015,Herath2015,Payer2016}.
This countermeasure is based on the assumption that all cache attacks cause significantly more cache hits and cache misses than benign applications.
While this assumption seems reasonable, it is unknown whether there are cache attacks that do not cause a significant number of cache hits and cache misses.

In this article, we present the \FlushOnly attack. 
\FlushOnly exploits the fact that the execution time of the \clflush instruction is shorter if the data is not cached and higher if the data is cached. At the same time, the \clflush instruction evicts the corresponding data from all cache levels. 
\FlushOnly exploits the same hardware and software properties as \FlushReload~\cite{Yarom2014}: it works on read-only shared memory, cross-core attack and in virtualized environments.
In contrast to \FlushReload, \FlushOnly does not make any memory accesses and thus does not cause any cache misses at all and only a minimal number of cache hits.
This distinguishes \FlushOnly from any other cache attack.
However, with both \FlushReload and \FlushOnly the victim process experiences an increased number of cache misses.

We evaluate \FlushOnly both in terms of \textit{performance} and \textit{detectability} in three scenarios: a covert channel, a side-channel attack on user input, and a side-channel attack on AES with T-tables.
We implement a detection mechanism that monitors cache references and cache misses of the last-level cache, similarly to state of the art~\cite{Chiappetta2015,Herath2015,Payer2016}. We show that existing cache attacks as well as Rowhammer attacks can be detected using performance counters.
However, we demonstrate that this countermeasure is non-effective against the \FlushOnly attack, as the fundamental assumption fails. The \FlushOnly attack is thus more stealthy than existing cache attacks, \ie a \FlushOnly spy process cannot be detected based on cache hits and cache misses. Thus, it cannot be detected by state-of-the-art detection mechanisms.

The \FlushOnly attack runs in a higher frequency and thus is faster than any existing cache attack in side-channel and covert channel scenarios. It achieves a cross-core transmission rate of 496\,KB/s, which is $6.7$ times faster than any previously published cache covert channel. The \FlushOnly attack does not trigger prefetches and thus allows to monitor multiple addresses within a 4\,KB memory range in contrast to \FlushReload that fails in these scenarios~\cite{Gruss2015Template}.

Our key contributions are:
\begin{itemize}

	\item We detail a new cache attack technique that we call \FlushOnly. It relies only on the difference in timing of the \clflush instruction between cached and non-cached memory accesses.
	\item We show that in contrast to all other attacks, \FlushOnly is stealthy, \ie it cannot be detected using hardware performance counters. We show that \FlushOnly also outperforms all existing cache attacks in terms of speed.
\end{itemize}

The remainder of this paper is organized as follows. Section~\ref{sec:background} provides background information on CPU caches, shared memory, and cache attacks.
Section~\ref{sec:flushonly} describes the \FlushOnly attack.
Section~\ref{sec:pmc} investigates how to leverage hardware performance counters to detect cache attacks.
We compare the performance and detectability of \FlushOnly attacks compared to state-of-the-art attacks in three scenarios: a covert channel in Section~\ref{sec:covertchannels}, a side-channel attack on keystroke timings in Section~\ref{sec:keystrokes}, and on cryptographic algorithms in Section~\ref{sec:crypto}.
Section~\ref{sec:discussion} discusses implications and countermeasures.
Section~\ref{sec:related} discusses related work. 
Finally, we conclude in Section~\ref{sec:conclusion}.

\section{Background}\label{sec:background}

\subsection{CPU Caches}\label{sec:caches}
CPU caches hide the memory accesses latency to the slow physical memory by buffering frequently used data in a small and fast memory. Modern CPU architectures implement $n$-way set-associative caches, where the cache is divided into cache sets, and each cache set comprises several cache lines. A line is loaded in a set depending on its address, and each line can occupy any of the $n$ ways.

On modern Intel processors, there are three cache levels.
The L3 cache, also called last-level cache, is shared between all CPU cores. The L3 cache is inclusive, \ie all data within the L1 and L2 caches is also present in the L3 cache.
Due to these properties, executing code or accessing data on one core has immediate consequences even for the private caches of the other cores. This can be exploited in so called cache attacks.
The last-level cache is divided into as many slices as cores, interconnected by a ring bus. Since the Sandy Bridge microarchitecture, each physical address is mapped to a slice by an undocumented so-called \textit{complex-addressing} function, that has recently been reversed-engineered~\cite{Maurice2015RAID}.

A cache replacement policy decides which cache line to replace when loading new data in a set. Typical replacement policies are least-recently used (LRU), variants of LRU and bimodal insertion policy where the CPU can switch between the two strategies to achieve optimal cache usage~\cite{Qureshi2007}. The unprivileged \clflush instruction evicts a cache line from all the cache hierarchy. However, a program can also evict a cache line by accessing enough memory.

\subsection{Shared Memory}\label{sec:shm}
Operating systems and hypervisors instrument shared memory to reduce the overall physical memory utilization and the TLB utilization. Shared libraries are loaded into physical memory only once and shared by all programs using them.
Thus, multiple programs access the same physical pages mapped within their own virtual address space.

The operating system similarly optimizes mapping of files, forking a process, starting a process twice, or using \verb|mmap| or \verb|dlopen|.
All cases result in a memory region shared with all other processes mapping the same file.

On personal computers, smartphones, private cloud systems and even in public clouds~\cite{Barresi2015}, another form of shared memory can be found, namely content-based page deduplication. The hypervisor or operating system scans the physical memory for byte-wise identical pages. Identical pages are remapped to the same physical page, while the other page is marked as free.
This technique can lower the use of physical memory and TLB significantly.
However, sharing memory between completely unrelated and possibly sandboxed processes, and between processes running in different virtual machines brings up security and privacy concerns.

\subsection{Cache Attacks and Rowhammer}\label{sec:cacheattacks}
Cache attacks exploit timing differences caused by the lower latency of CPU caches compared to physical memory. Access-driven cache attacks are typically devised in two types: \PrimeProbe~\cite{Osvik2006,Percival2005,Tromer2010} and \FlushReload~\cite{Gullasch2011,Yarom2014}.

In \PrimeProbe attacks, the attacker occupies a cache set and measures whenever a victim replaces a line in that cache set.
Modern processors have a physically indexed last-level cache, use complex addressing, and undocumented replacement policies. Cross-VM side-channel attacks~\cite{Irazoqui2015SA,Liu2015} and covert channels~\cite{Maurice2015C5} that tackle these challenges have been presented in the last year.
Oren~\etal\cite{Oren2015} showed that a \PrimeProbe cache attack can be launched from within sandboxed JavaScript in a browser, allowing a remote attacker to eavesdrop on network traffic statistics or mouse movements through a website.

\FlushReload is a two phase attack that works on a single cache line. First, it \textit{flushes} a cache line using the \clflush instruction, then it measures the time it takes to \textit{reload} the data. Based on the time  measurement, the attacker determines whether a targeted address has been reloaded by another process in the meantime.
In contrast to \PrimeProbe, \FlushReload exploits the availability of shared memory and especially shared libraries between the attacker and the victim program.
Applications of \FlushReload have been shown to be reliable and powerful, mainly to attack cryptographic algorithms~\cite{Irazoqui2015Neighbor,Irazoqui2015Lucky,Guelmezoglu2015,Zhang2014}. 

Rowhammer is not a typical cache attack but a DRAM vulnerability that causes random bit flips by repeatedly accessing a DRAM row~\cite{Kim2014}. It however shares some similarities with caches attacks since the accesses must bypass all levels of caches to reach DRAM and trigger bit flips. Attacks exploiting this vulnerability have already been demonstrated to gain root privileges and to evade a sandbox~\cite{SeabornBlackhat2015}.
Rowhammer causes a significant number of cache hits and cache misses, that resemble a cache attack.

\section{The \emph{\flushonly} Attack}\label{sec:flushonly}

The \FlushOnly attack is a faster and stealthier alternative to existing cache attacks that also has fewer side effects on the cache. In contrast to other cache attacks, it does not perform any memory accesses. For this reason it causes no cache misses and only a minimal number of cache hits. Thus, proposed detection mechanisms based on hardware performance counters fail to detect the \FlushOnly attack. \FlushOnly exploits the same hardware and software properties as \FlushReload. It runs across cores and in virtualized environments if read-only shared memory with the victim process can be acquired.

Our attack builds upon the observation that the \clflush instruction can abort early in case of a cache miss. In case of a cache hit, it has to trigger eviction on all local caches. This timing difference can be exploited in form of a cache attack, but it can also be used to derive information on cache slices and CPU cores as each core can access its own cache slice faster than others.

The attack consists of only one phase, that is executed in an endless loop. It is the execution of the \clflush instruction on a targeted shared memory line.
The attacker measures the execution time of the \clflush instruction. Based on the execution time, the attacker decides whether the memory line has been cached or not. As the attacker does not load the memory line into the cache, this reveals whether some other process has loaded it. At the same time, \clflush evicts the memory line from the cache for the next loop round of the attack.

The measurement is done using the \texttt{rdtsc} instruction that provides a sub-nanosecond resolution timestamp. It also uses \texttt{mfence} instructions, as \clflush is only ordered by \texttt{mfence}, but not by any other means. 

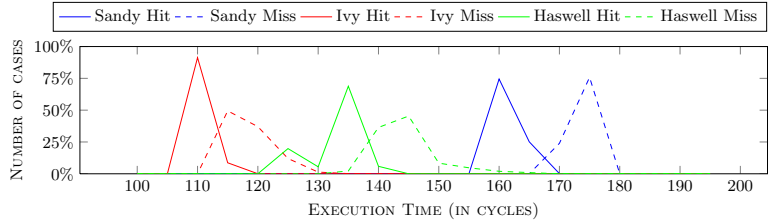
\begin{figure}
\centering
\begin{tikzpicture}[scale=0.7]
\pgfplotsset{every axis legend/.append style={at={(0.48,1.35)},anchor=north}}
\begin{axis}[
legend columns=6,
xlabel=\textsc{Execution Time (in cycles)},
ylabel=\textsc{Number of cases},
ymin=0,
ymax=1,
ytick={0,0.25,0.5,0.75,1},
yticklabels={$0\%$,$25\%$,$50\%$,$75\%$,$100\%$},
width=1.2\hsize,
height=4cm
]
\addplot+[no marks,blue,draw=blue,mark=*,mark options={draw=blue,fill=blue}] table[x=t,y=SandyHit] {histogram_comparison.csv};
\addlegendentry{Sandy Hit}
\addplot+[no marks,blue,dashed,draw=blue,mark=square*,mark options={draw=blue,fill=blue}] table[x=t,y=SandyMiss] {histogram_comparison.csv};
\addlegendentry{Sandy Miss}
\addplot+[no marks,red,draw=red,mark=*,mark options={draw=red,fill=red}] table[x=t,y=IvyHit] {histogram_comparison.csv};
\addlegendentry{Ivy Hit}
\addplot+[no marks,red,dashed,draw=red,mark=square*,mark options={draw=red,fill=red}] table[x=t,y=IvyMiss] {histogram_comparison.csv};
\addlegendentry{Ivy Miss}
\addplot+[no marks,green,draw=green,mark=*,mark options={draw=green,fill=green}] table[x=t,y=HaswellHit] {histogram_comparison.csv};
\addlegendentry{Haswell Hit}
\addplot+[no marks,green,dashed,draw=green,mark=square*,mark options={draw=green,fill=green}] table[x=t,y=HaswellMiss] {histogram_comparison.csv};
\addlegendentry{Haswell Miss}
\end{axis}
\end{tikzpicture}
\caption{Execution time of the \clflush instruction on cached and uncached memory on different CPU architectures}
\label{fig:histogram}
\end{figure}

Figure~\ref{fig:histogram} shows the execution time histogram of the \clflush instruction for cached and non-cached memory lines, run on the three setups with different recent microarchitectures: a Sandy Bridge i5-2540M, an Ivy Bridge i5-3320M and a Haswell i7-4790. The timing difference of the peaks is 12 cycles on Sandy Bridge, 9 cycles on Ivy Bridge, and 12 cycles on Haswell. If the address maps to a remote core, another penalty of 3 cycles is added to the minimum execution time for cache hits. The difference is enough to be observed by an attacker. We discuss this timing difference and its implications in Section~\ref{sec:related_clflush}. In either case the execution time is less than the access time for both memory cached in the last-level cache and memory accesses that are not cached. Therefore, \FlushOnly is significantly faster than any other last-level cache attack.

The \FlushOnly attack inherently has a slightly lower accuracy than the \FlushReload technique in some cases, due to the lower timing difference between a hit and a miss and because of a lower access time on average. Nevertheless, the same amount of information is extracted faster using the \FlushOnly attack due to the significantly lower execution time. Furthermore, the reload-step of the \FlushReload attack can trigger the prefetcher and thus destroy measurements by fetching data into the cache. This is the case especially when monitoring more than one address within a physical page~\cite{Gruss2015Template}. As the \FlushOnly attack never performs any memory accesses, this problem does not exist and the \FlushOnly attack achieves an even higher accuracy here. For the same reason, the \FlushOnly attack causes no cache misses and only a minimal number of cache hits.
Thus, recently proposed detection mechanisms using cache references and cache misses fail to detect \FlushOnly.

\section{Detecting Cache Attacks with Hardware Performance Counters}\label{sec:pmc}
Cache attacks can lead to an increased number of cache hits or cache misses in the attacker process or in other processes. Thus, it may be possible to detect abnormal behavior on a system level.
However, to stop or prevent an attack, it is necessary to identify the attacking process.
Therefore, we consider an attack \textit{stealthy} if the attacking spy process cannot be identified.

Hardware performance counters are special-purpose registers that are used to monitor special hardware-related events. Events that can be monitored include cache references and cache misses on the last-level cache. 
They are mostly used for performance analysis and fine tuning, but have been found to be suitable to detect Rowhammer and the \FlushReload attack~\cite{Herath2015,Chiappetta2015,Payer2016}. The focus of our work is to show that detection of existing attacks is straight-forward, but detection of the \FlushOnly attack using these performance counters is infeasible, due to the absence of cache misses and the minimal number of cache references.

We analyze the feasibility of such detection mechanisms using the Linux \verb|perf_event_open| syscall interface that provides userspace access to a subset of all available performance counters on a per-process basis. The actual accesses to the model specific registers are performed in the kernel. The same information can be used by a system service to detect ongoing attacks. During our tests we ran the performance monitoring with system service privileges.

\begin{table}[t!]
\centering
\caption{List of hardware performance events we use.} \label{tab:performance_events}
\resizebox{\hsize}{!}{
\csvautobookptabular[respect underscore=true,separator=semicolon]{performance_events.csv} 
}
\end{table}

We analyzed all 23 hardware and cache performance events available with the Linux syscall interface on our system. Additionally, we analyzed the so called \textit{uncore}~\cite{Intel_vol3} performance monitoring units and found one called C-Box that is influenced by cache hits, misses and \clflush instructions directly.
The \texttt{UNC\_CBO\_CACHE\_LOOKUP} event of the C-Box allows monitoring a last-level cache lookups per cache slice, including by the \texttt{clflush} instruction. The C-Box monitoring units are not available through a generic interface but only through model specific registers. 
Table~\ref{tab:performance_events} lists all events we evaluated.
We found that there are no other performance counters documented to monitor cache hits, misses or \clflush instructions specifically. Furthermore, neither the hypervisor nor the operating system can intercept the \clflush instruction or monitor the frequency of \clflush instructions being executed using performance counters.

The number of performance events that can be monitored simultaneously is limited by hardware. On all our test systems it is possible to monitor up to 4 events simultaneously. Thus, any detection mechanism can only use 4 performance events simultaneously.

We evaluated the 24 performance counters for the following scenarios:
\begin{enumerate}
\item Idle: idle system,
\item Firefox: user scrolling down a chosen Twitter feed in Firefox,
\item OpenTTD: user playing a game
\item stress -m 1: loop reading and writing in dynamically allocated 256\,MB arrays,
\item stress -c 1: loop doing a CPU computation with almost no memory,
\item stress -i 1: loop calling the I/O \texttt{sync()} function,
\item \FlushReload: cache attack on the GTK library to spy on keystroke events,
\item Rowhammer: Rowhammer attack.
\end{enumerate}
The first 3 scenarios are casual computer usage scenarios, the next 3 cause a benign high load situation and the last 2 perform an attack.
A good detection mechanism classifies as benign the scenarios 1 to 6 and as attacks 7 and 8.

We use the instruction TLB (ITLB) performance counters ($\texttt{ITLB\_RA} + \texttt{ITLB\_WA}$) to normalize the performance counters to make cache attacks easier to detect, and prevent scenarios 2 and 3 from being detected as malicious. 
Indeed, the main loop that is used in the \FlushReload and Rowhammer attacks causes a high number of last-level cache misses while executing only a small piece of code. Executing only a small piece of code causes a low pressure on the ITLB. 

\begin{table}[t]
\centering
\caption{Comparison of performance counters normalized to the number of ITLB events in different cache attacks and normal scenarios over 135 seconds in separate runs.} \label{tab:comparison_references_misses_1}

\newcolumntype{x}{r}
\resizebox{\hsize}{!}{
\csvautobookpxtabular[respect percent=true,separator=semicolon]{comparison_references_misses_1.csv}
}
\end{table}

Table~\ref{tab:comparison_references_misses_1} shows a comparison of performance counters for the 8 scenarios tested over 135 seconds. These tests were performed in multiple separate runs as the performance monitoring unit can only monitor 4 events simultaneously.
Not all cache events are suitable for detection. The \verb|UNC_CBO_CACHE_LOOKUP| event that counts cache slice events including \texttt{clflush} operations shows very high values in case of \texttt{stress -i}. It would thus lead to false positives. 
Similarly, the \verb|INSTRUCTIONS| event used by Chiappetta~\etal\cite{Chiappetta2015} has a significantly higher value in case of \texttt{stress -c} than in the attack scenarios and would cause false positives in the case of benign CPU intensive activities.
The \verb|REF_CPU_CYCLES| is the unscaled total number of CPU cycles consumed by the process. Divided by the TLB events, it shows how small the executed loop is. The probability of false positive matches is high, for instance in the case of \texttt{stress -c}.

Thus, 4 out of 24 events allow detecting both \FlushReload and Rowhammer without causing false positives for benign applications. 
The rationale behind these events is as follows:
\begin{enumerate}
\item \verb|CACHE_MISSES| occur after data has been flushed from the last-level cache,
\item \verb|CACHE_REFERENCES| occur when reaccessing memory,
\item \verb|L1D_RM| occur because flushing from last-level cache also flushes from the lower cache levels,
\item \verb|LL_RA| are a subset of the \verb|CACHE_REFERENCES| counter, they occur when reaccessing memory,
\end{enumerate}
Two of the events are redundant: \verb|L1D_RM| is redundant with \verb|CACHE_MISSES|, and \verb|LL_RA| with \verb|CACHE_REFERENCES|.
We will thus focus on the \verb|CACHE_MISSES| and \verb|CACHE_REFERENCES| events as proposed in previous work~\cite{Herath2015,Chiappetta2015,Payer2016}.

We define that a process is considered malicious if more than $k_m$ cache miss or $k_r$ cache reference per ITLB event are observed. The attack is detected if
\[\frac{C_{\texttt{CACHE\_MISSES}}}{C_{\texttt{ITLB\_RA}}+C_{\texttt{ITLB\_WA}}} \geq k_m,\text{ \, \, or \, \, }\frac{C_{\texttt{CACHE\_REFERENCES}}}{C_{\texttt{ITLB\_RA}}+C_{\texttt{ITLB\_WA}}} \geq k_r,\]
with $C$ the value of the corresponding performance counter. The operating system can choose the frequency in which to run the detection checks.

The thresholds for the cache reference and cache hit rate are determined based on a set of benign applications and malicious applications. It is chosen to have the maximum distance to the minimum value for any malicious application and the maximum value for any benign application. In our case this is $k_m=2.35$ and $k_r=2.34$. Based on these thresholds, we perform a classification of processes into malicious and benign processes. We tested this detection mechanism against various cache attacks and found that it is suitable to detect different \FlushReload, \PrimeProbe and Rowhammer attacks as malicious.
However, the focus of our work is not the evaluation of detection mechanisms based on performance counters, but to show that such detection mechanisms cannot reliably detect the \FlushOnly attack due to the absence of cache misses and a minimal number of cache references.

In the following sections, we evaluate the performance and the detectability of \FlushOnly compared to the state-of-the-art cache attacks \FlushReload and \PrimeProbe in three scenarios: a covert channel, a side channel on user input and a side channel on AES with T-tables.

\section{Covert Channel Comparison}\label{sec:covertchannels}
In this section, we describe a generic low-error cache covert channel framework. In a covert channel, an attacker runs two unprivileged applications on the system under attack. The processes are cooperating to communicate with each other, even though they are not allowed to by the security policy.
We show how the two processes can communicate using the \FlushOnly, \FlushReload, and \PrimeProbe technique. We compare the performance and the detectability of the three implementations. In the remainder of the paper, all the experiments are performed on a Haswell i7-4790 CPU.

\subsection{A Low-error Cache Covert Channel Framework}
In order to perform meaningful experiments and obtain comparable and fair results, the experiments must be reproducible and tested in the same conditions. This includes the same hardware setup, and the same protocols.
Indeed, we cannot compare covert channels from published work~\cite{Maurice2015C5,Liu2015} that have different capacities and error rates. Therefore, we build a framework to evaluate covert channels in a reproducible way. This framework is generic and can be implemented over any covert channel that allows bidirectional communication, by implementing the \verb|send()| and \verb|receive()| functions.

The central component of the framework is a simple transmission protocol.
Data is transmitted in packets of $N$ bytes, consisting of $N-3$ bytes payload, a $1$ byte sequence number and a CRC-16 checksum over the packet. The sequence number is used to distinguish consecutive packets. The sender retransmits packets until the receiver acknowledges it. Packets are acknowledged by the receiver if the checksum is valid.

Although errors are still possible in case of a false positive CRC-16 checksum match, the probability is low. We choose the parameters such that the effective error rate is below $5\%$. The channel capacity measured with this protocol is comparable and reproducible. Furthermore, it is close to the effective capacity in a real-world scenario, because error-correction cannot be omitted.
The number of transmitted bits is the minimum of bits sent and bits received. The transmission rate can be computed by dividing the number of transmitted bits by the runtime. The error rate is given by the number of all bit errors between the sent bits and received bits, divided by the number of transmitted bits.

\subsection{Covert Channel Implementations}
We first implemented the \FlushReload covert channel. By accessing fixed memory locations in a shared library the a 1 is transmitted, whereas a 0 is transmitted by omitting the access.
The receiver performs the actual \FlushReload attack to determine whether a 1 or a 0 was transmitted.
The bits retrieved are then parsed as a data frame according to the transmission protocol. The sender also monitors some memory locations using \FlushReload for cache hits too, to receive packet acknowledgments.

\begin{figure}
\centering
\includegraphics[width=0.5\hsize]{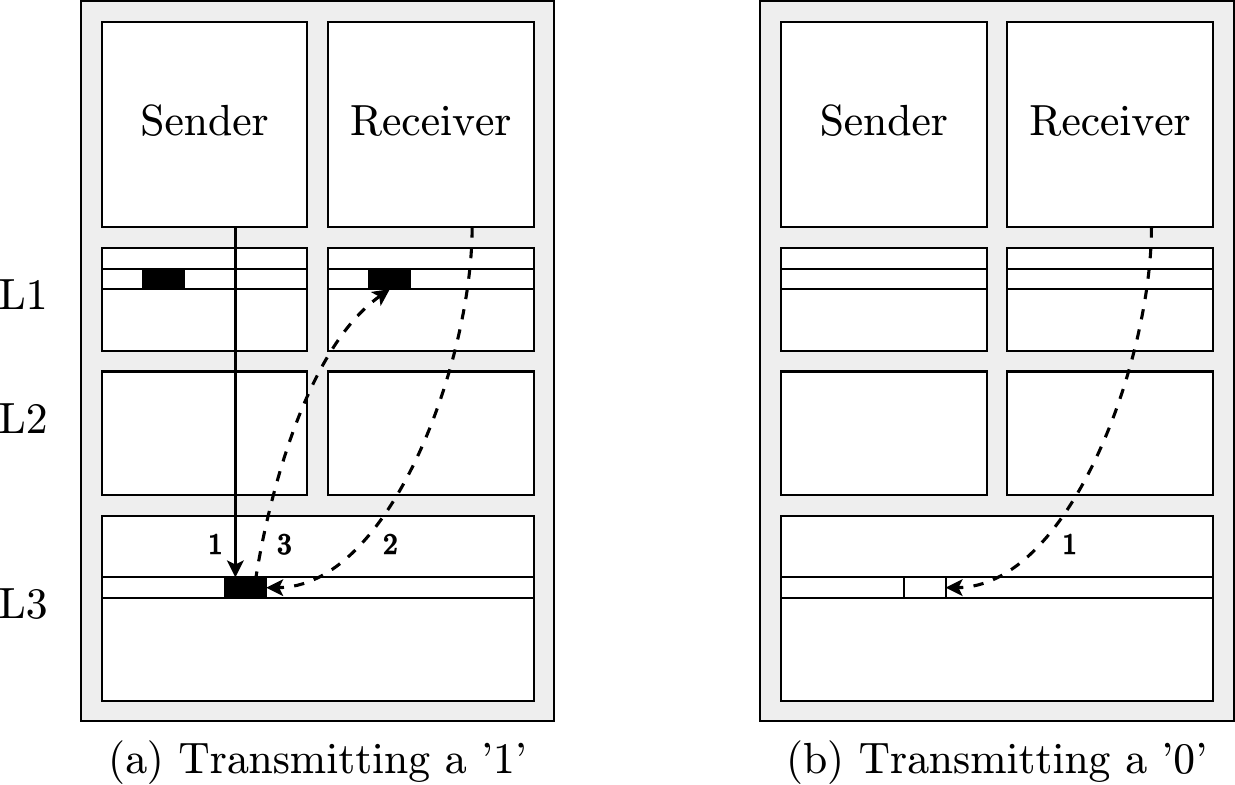}
\caption{Illustration of the \FlushOnly covert channel.}\label{fig:flushonly}
\end{figure}

The second implementation is the \FlushOnly covert channel, illustrated by Figure~\ref{fig:flushonly}. It uses the same sender process as the \FlushReload covert channel. 
To transmit a 1 (Figure~\ref{fig:flushonly}-a), the sender accesses the memory location, that is cached (step 1). This time, the receiver only flushes the shared line. As the line is present in the last-level cache by inclusiveness, it is flushed from this level (step 2). A bit also indicates that the line is present in the L1 cache, and thus must also be flushed from this level (step 3).
To transmit a 0 (Figure~\ref{fig:flushonly}-b), the sender stays idle. The receiver flushes the line (step 1). As the line is not present in the last-level cache, it means that it is also not present in the lower levels, which results in a faster execution of the \texttt{clflush} instruction.
Thus only the sender process performs memory accesses, while the receiver only flushes cache lines. 
To send acknowledgment bytes the receiver performs memory accesses and the sender runs a \FlushOnly attack.

The third implementation is the \PrimeProbe covert channel. It uses the same attack technique as Liu~\etal\cite{Liu2015}, Oren~\etal\cite{Oren2015}, and Maurice~\etal\cite{Maurice2015C5}.
The sender transmits a 1 bit by priming a cache set. The receiver probes the same cache set. Again the receiver determines whether a 1 or a 0 was transmitted.
We make two adjustments for convenience and to focus solely on the transmission part.
First, we compute a static eviction set by using the complex addressing function~\cite{Maurice2015RAID} on physical addresses. This avoids the possibility of errors introduced by timing-based eviction set computation. 
Second, we map the shared library into our address space to determine the physical address to attack to make an agreement on the cache sets in sender and receiver. Yet, the shared library is never accessed and unmapped even before the \PrimeProbe attack is started.
We assume that the sender and receiver have agreed on the cache sets in a preprocessing step. This is practical even for a timing-based approach.

\subsection{Performance Evaluation}

Table~\ref{tab:covert_all_in_one} compares the capacity and the detectability of the three covert channels in different configurations. The \FlushOnly covert channel is the fastest of the three covert channels. With a packet size of 28 bytes the transmission rate is 496\,KB/s. At the same time the effective error rate is only $0.84\%$.
The \FlushReload covert channel also achieved a good performance at a packet size of 28 bytes. The transmission rate then is 298\,KB/s and the error rate $<0.005\%$. With a packet size of 4 bytes, the performance is lower in all three cases.

\begin{table*}[t]
\centering
\caption{Comparison of capacity and detectability of the three cache covert channels with different parameters. \FlushOnly and \FlushReload use the same sender process.}\label{tab:covert_all_in_one}

\newcolumntype{x}[1]{
>{\raggedleft}p{#1}}

\newcolumntype{y}[1]{
>{\raggedleft}p{#1}}

\csvautobookxxtabular[separator=semicolon]{covert_all_in_one.csv}
\end{table*}

A \PrimeProbe covert channel with a 28-byte packet size is not realistic. First, to avoid triggering the hardware prefetcher we do not access more than one address per physical page. Second, for each eviction set we need 16 addresses. Thus we would require $28\textnormal{B} \cdot 4096 \cdot 16 = 14\,\textnormal{GB}$ of memory only for the eviction sets.
For \PrimeProbe we achieved the best results with a packet size of 5 bytes. With this configuration the transmission rate is 68\,KB/s at an error rate of $0.14\%$, compared to 132\,KB/s using \FlushReload and 95\,KB/s using \FlushOnly.

The \FlushOnly transmission rate of 496\,KB/s is significantly higher than any other state-of-the-art cache covert channels. It is $6.7$ times as fast as the fastest cache covert channel to date~\cite{Liu2015} at a comparable error rate. Our covert channel based on \FlushReload is also faster than previously published cache covert channels, but still much slower than the \FlushOnly covert channel. Compared to our \PrimeProbe covert channel, \FlushOnly is $7.3$ times faster.

\subsection{Detectability}

Table~\ref{tab:covert_all_in_one} shows the evaluation of the detectability for packet sizes that yielded the highest performance in one of the cases.
\FlushReload and \FlushOnly use the same sender process, the reference and miss count is mainly influenced by the number of retransmissions and executed program logic.
\FlushReload is detected in all cases either because of its sender or its receiver, although its sender process with a 4-byte packet size stays below the detection threshold.
The \PrimeProbe attack is always well above the detection threshold and therefore always detected as malicious.
All \FlushOnly receiver processes are classified as benign.
However, only the sender process used for the \FlushOnly and the \FlushReload covert channels with a 4-byte packet size is classified as benign.

The receiver process performs most of the actual cache attack. If it is sufficient to keep the receiver process stealthy, \FlushOnly clearly outperforms all other cache attacks.
If the sender has to be stealthy as well, the sender process used by \FlushOnly and \FlushReload performs better than the \PrimeProbe sender process.
However, due to the high number of cache hits it is difficult to keep the sender process below the detection threshold.
An adversary could choose to reduce the transmission rate in order to be stealthier in either case.

\section{Side-Channel Attack on User Input}\label{sec:keystrokes}
Another cache attack that has been demonstrated recently using \FlushReload, is eavesdropping on keystroke
timings. We attack an address in the GTK library invoked when processing keystrokes.
The attack is implemented as a program that constantly flushes the address, and derives when a keystroke occurred, based on memory access times or the execution time of the \clflush instruction.

\subsection{Performance Evaluation}

We compare the three attacks \FlushOnly, \FlushReload, and \PrimeProbe, based on their performance in this side-channel attack scenario.
During each test we simulate a user typing a 1000-character text into an editor. Each test takes 135 seconds.
As expected, \FlushReload has a very high accuracy of $96.1\%$. This allows direct logging of keystroke timings. \FlushOnly performs notably well, with $74.7\%$ correctly detected keystrokes. However, this makes a practical attack much harder than with \FlushReload. The attack with \PrimeProbe yielded no meaningful results at all due to the high noise level.
In case of \FlushReload and \FlushOnly the accuracy can be increased significantly by attacking 3 addresses that are used during keystroke processing simultaneously. The decision whether a keystroke was observed is then based on these 3 addresses increasing the accuracy significantly. Using this technique reduces the error rate in case of \FlushReload close to $100\%$ and above $92\%$ in case of \FlushOnly.

\subsection{Detectability}
\floatsetup[table]{font=small}
\begin{table}[t]
\centering
\caption{Comparison of performance counters normalized to the number of ITLB events for cache attacks on user input.}\label{tab:comparison_sc_performance_counter}
\begin{tabular}{lrrr}
\toprule
Technique & \quad Cache references & \quad Cache misses & \quad Stealthy \\
\midrule
{\FlushReload} & $5.140$ & $5.138$ & \xmark \\
\textbf{\FlushOnly} & $0.002$ & $0.000$ & \cmark \\
\bottomrule
\end{tabular}
\end{table}

To evaluate the detectability we again monitored the cache references and cache misses events, and compared the three cache attacks with each other and with an idle system.
Table~\ref{tab:comparison_sc_performance_counter} shows that \FlushReload generates a high number of cache references, whereas \FlushOnly causes a negligible number of cache references. We omitted \PrimeProbe in this table as it was not sufficiently accurate to perform the attack.

\FlushReload yields the highest accuracy in this side-channel attack, but it is easily detected. The accuracy of \FlushOnly can easily be increased to more than $92\%$ and it still is far from being detected. Thus, \FlushOnly is a viable and stealthy alternative to the \FlushReload attack as it is not classified as malicious based on the cache references or cache misses performance counters.

\section{Side-Channel Attack on AES with T-Tables}\label{sec:crypto}
To round up our comparison with other cache attacks, we compare \FlushOnly, \FlushReload, and \PrimeProbe in a high frequency side-channel attack scenario.
Finding new cache attacks is out of scope of our work. Instead, we try to perform a fair comparison between the different attack techniques by implementing a well known cache attack using the three techniques on a vulnerable implementation of a cryptographic algorithm.
We attack the OpenSSL T-Table-based AES implementation that is known to be susceptible to cache attacks~\cite{Bernstein2005,Osvik2006}. 
This AES implementation is disabled by default for security reasons, but still exists for the purpose of comparing new and existing side-channel attacks.

The AES algorithm uses the T-tables to compute the ciphertext based on the secret key $k$ and the plaintext $p$. During the first round, table accesses are made to entries $T_j[p_i \oplus k_i]$ with $i \equiv j \mod 4$ and $0 \leq i < 16$. Using a cache attack it is possible to derive values for $p_i \oplus k_i$ and thus, possible key-byte values $k_i$ in case $p_i$ is known.

\subsection{Attack Implementation Using \flushonly}
The implementation of the chosen-plaintext attack  side-channel attacks for the three attack techniques is very similar. The attacker triggers an encryption, choosing $p_i$ while all $p_j$ with $i\neq j$ are random. One cache line holds 16 T-Table entries. The cache attack is now performed on the first line of each T-Table. The attacker repeats the encryptions with new random plaintext bytes $p_j$ until only one $p_i$ remains to always cause a cache hit. The attacker learns that $p_i \oplus k_i \equiv_{\mceil{4}} 0$ and thus $k_i \equiv_{\mceil{4}} p_i$.
After performing the attack for all 16 key bytes, the attacker has derived 64 bits of the secret key $k$. As we only want to compare the three attack techniques, we do not extend this attack to a full key recovery attack.

\subsection{Performance Evaluation}

\begin{figure}[t]
\centering
\includegraphics[width=0.15\hsize]{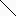} \quad
\includegraphics[width=0.15\hsize]{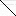} \quad
\includegraphics[width=0.15\hsize]{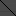}
\caption{Comparison of Cache Templates (address range of the first T-table) generated using \FlushReload (left), \FlushOnly (middle), and \PrimeProbe (right). In all cases $k_0=\mathtt{0x00}$.}\label{opensslscan}
\end{figure}

\floatsetup[table]{font=small}
\begin{table}[t]
\centering
\caption{Number of encryptions to determine the upper 4 bits of a key byte.}\label{tab:number_encryptions}
\begin{tabular}{lr}
\toprule
Technique & \quad Number of encryptions \\
\midrule
\FlushReload & 250 \\
\textbf{\FlushOnly} & 350 \\
\PrimeProbe & 4\,800 \\
\bottomrule
\end{tabular}
\end{table}

Figure~\ref{opensslscan} shows a comparison of cache templates generated with \FlushReload, \FlushOnly, and \PrimeProbe using $1\,000\,000$
encryptions to create a visible pattern in all three cases. Similar templates can be found in previous work~\cite{Osvik2006,Spreitzer2013COSADE,Gruss2015Template}. Table~\ref{tab:number_encryptions} shows how many encryptions are necessary to determine the upper 4 bits correctly. We performed encryptions until the correct guess for the upper 4 bits of key byte $k_0$ had a $5\%$ margin over all other key candidates. \FlushOnly requires around $1.4$ times as many encryptions as \FlushReload, but $13.7$ times less than \PrimeProbe to achieve the same accuracy.

\FlushOnly is the only attack that does not trigger the prefetcher. Thus, we can monitor multiple adjacent cache sets. By doing this we double the number of cache references, but increase the accuracy of the measurements so that 275 encryptions are sufficient to identify the correct key byte with a $5\%$ margin. That is only $1.1$ times as many encryptions as \FlushReload and $17.5$ times less than \PrimeProbe. Thus, \FlushOnly on multiple addresses is faster at deriving the same information as \FlushReload.

\subsection{Detectability}

\begin{table*}[t]
\centering
\caption{Comparison of the performance counters when performing 256 million encryptions with different cache attacks and without an attack.}\label{tab:comparison_aes_performance_counter}

\newcolumntype{x}[1]{
>{\raggedleft}p{#1\hsize}}

\begin{tabular}{lx{0.15}x{0.15}x{0.13}x{0.14}x{0.11}r}
\toprule
Technique & Cache references & Cache misses & Execution time in s & References (norm.) & Misses (norm.) & \, Stealthy \\
\midrule
{\FlushReload} & $1\,024 \cdot 10^6$ & 19\,284\,602 & 215 & 2\,513.43 & 47.33 & \xmark \\
{\PrimeProbe} & $4\,222 \cdot 10^6$ & 294\,897\,508 & 234 & 1\,099.63 & 76.81 & \xmark \\
{\textbf{\FlushOnly}} & $768 \cdot 10^6$ & 1\,741 & 163 & 1.40 & 0.00 & \cmark \\
\bottomrule
\end{tabular}
\end{table*}

Table~\ref{tab:comparison_aes_performance_counter} shows a comparison of the performance counters for the three attacks over 256 million encryptions. The \FlushOnly attack took only 163 seconds whereas \FlushReload took 215 seconds and \PrimeProbe 234 seconds for the identical attack.
On a system level, it is possible to notice ongoing cache attacks on AES in all three cases due to the high number of cache misses caused by the AES encryption process. However, to stop or prevent the attack, it is necessary to detect the spy process.
\PrimeProbe exceeds the detection threshold by a factor of 468 and \FlushReload exceeds the threshold by a factor of 1070. To stay below the detection threshold, slowing down the attack by at least the same factor would be necessary. In contrast, \FlushOnly is not detected based on our classifier and does not have to be slowed down to be stealthy.

\section{Discussion}\label{sec:discussion}
\subsection{Using \clflush to Detect Cores and Cache Slices}

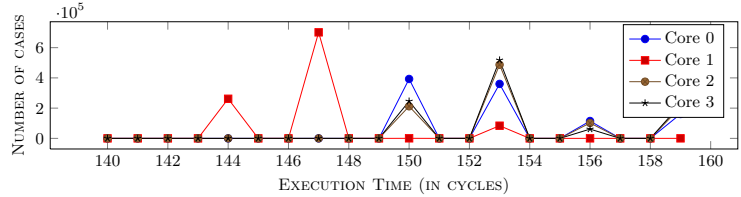
\begin{figure}
\centering
\begin{tikzpicture}[scale=0.7]
\begin{axis}[
xlabel=\textsc{Execution Time (in cycles)},
ylabel=\textsc{Number of cases},
width=1.2\hsize,
height=4cm
]
\addplot+ table[x=x,y=y] {histogram1.csv};
\addlegendentry{Core 0}
\addplot+ table[x=x,y=y] {histogram4.csv};
\addlegendentry{Core 1}
\addplot+ table[x=x,y=y] {histogram10.csv};
\addlegendentry{Core 2}
\addplot+ table[x=x,y=y] {histogram40.csv};
\addlegendentry{Core 3}
\end{axis}
\end{tikzpicture}
\caption{Excerpt of the \texttt{clflush} histogram for an address in slice 1 on different cores. The lower execution time on core 1 shows that this address maps to slice 1.}
\label{fig:slice_detection}
\end{figure}	
		
The \FlushOnly attack can be used to determine on which CPU core a process is running or to which cache slice an address maps.
Indeed, a \clflush on a remote cache slice takes longer than a \clflush on a local cache slice, as shown in Figure~\ref{fig:slice_detection}. This is due to the ring bus architecture connecting remote slices. 
Knowing the physical address of a memory access on a local slice, we can then use the complex addressing function~\cite{Maurice2015RAID} to determine on which core the process runs. However, this would require high privileges. Yet, it is possible to determine to which slice an address maps without knowing the physical address by performing a timing attack. This can be done by an unprivileged process, as pinning a thread to a CPU core requires no privileges.

This can be exploited to detect colocation on the same CPU, CPU core or hyperthreading core in restricted environments even if the \texttt{cpuid} instructions is virtualized. It is more difficult to determine which CPU core a thread runs on based on memory access timings because of the influence of lower level caches. Such an attack has also not been demonstrated yet. The information on the executing CPU core can be used to enhance cache attacks and other attacks such as the Rowhammer attack~\cite{Kim2014,Gruss2015Row}.
Running \clflush on a local slice lowers the execution time of each Rowhammer loop round by a few cycles. The probability of bit flips increases as the execution time lowers, thus we can leverage the information whether an address maps to a local slice to improve this attack.

A similar timing difference also occurs upon memory accesses that are served from the local or a remote slice respectively. The reason again is the direct connection to the local cache slice while remote cache slices are connected via a ring bus.
However, as memory accesses will also be cached in lower level caches, it is more difficult to observe the timing difference without \clflush.
The \clflush instruction directly manipulates the last-level cache, thus lower level caches cannot hide the timing difference.
 
While the operating system can restrict access on information such as the CPU core the process is running on and the physical address mapping to make efficient cache attacks harder, it cannot restrict access to the \clflush instruction. Hence, the effect of such countermeasures is lower than expected.

\subsection{Countermeasures}\label{sec:countermeasures}
We suggest modifying the \clflush instruction to counter the wide range of attacks that it can be used for. The difference in the execution time of \clflush is 3 cycles depending on the cache slice and less than 12 cycles depending on whether it is a cache miss. In practice the \clflush instruction is used only in rare situations and not in a high frequency. Thus, a hypothetical performance advantage cannot justify introducing these exploitable timing differences. We propose making \clflush a constant-time instruction. This would prevent the \FlushOnly attack completely, as well as information leakage on cache slices and CPU cores.

\FlushOnly is the only cache attack that does not perform any memory accesses and thus causes no cache misses and only a minimal number of cache references. One theoretical way to detect our attack would be to monitor each load, \eg by timing, and to stop when detecting too many misses. However, this solution is currently not practical, as a software-based solution that monitors each load would cause a significant performance degradation. A similar hardware-based solution called \textit{informing loads} has been proposed by Kong~\etal\cite{Kong2009}, however it needs a change in the instruction set. 
Without hardware modifications it would be possible to enable the \texttt{rdtsc} instruction only in privileged mode as can be done using seccomp on Linux~\cite{Linux2008} since 2008. Fogh~\cite{Fogh2015Blog} proposed to simulate the \texttt{rdtsc} in an interrupt handler, degrading the accuracy of measurements far enough to make cache attacks significantly harder.

\FlushReload and \FlushOnly both require shared memory. If shared memory is not available, an attacker would have to resort to a technique that even works without shared memory such as \PrimeProbe.
Furthermore, making the \clflush instruction privileged would prevent \FlushReload and \FlushOnly as well. However, this would require changes in hardware and could not be implemented in commodity systems.

\section{Related work}\label{sec:related}

	\subsection{Detecting and Preventing Cache Attacks}\label{sec:related_clflush}

Zhang~\etal\cite{Zhang2011} proposed HomeAlone, a system-level solution that uses a \PrimeProbe covert channel to \textit{detect} the presence of a foe co-resident virtual machine. The system monitors random cache sets so that friendly virtual machines can continue to operate if they change their workload, and that foe virtual machines are either detected or forced to be silent.
Cache Template Attacks~\cite{Gruss2015Template} can be used to detect attacks on shared libraries and binaries as a user. However, such a permanent scan increases the system load and can only detect attacks in a small address range within a reasonable response time.

Herath and Fogh~\cite{Herath2015} proposed to monitor cache misses to detect \FlushReload attacks and Rowhammer. The system would slow down or halt all attacker processes. With the detection mechanism we implemented, we show that this technique is feasible for previous attacks but not for the \FlushOnly attack.
Chiappetta~\etal\cite{Chiappetta2015} proposed to build a trace of cache references and cache misses over the number of executed instructions to detect \FlushReload attacks.
They then proposed three methods to analyze this trace: a correlation-based method, and two other ones based on machine learning techniques.
However, a learning phase is needed to detect malicious programs that are either from a set of known malicious programs or resemble a program from this set. They are thus are less likely to detect new or unknown cache attacks or Rowhammer attacks, in contrast to our ad-hoc detection mechanism.
Payer~\cite{Payer2016} proposed a system called HexPADS to use cache references, cache misses, but also other events like page faults to detect cache attacks and Rowhammer at runtime.

Cache attacks can be \emph{prevented} at three levels: at the hardware level, at the system level, and finally, at the application level. 
At the hardware level, several solutions have been proposed to prevent cache attacks, either by removing cache interferences, or randomizing them. The solutions include new secure cache designs~\cite{Wang2007,Wang2008,Liu2014} or altering the prefetcher policy~\cite{Fuchs2015}. However, hardware changes are not applicable to commodity systems.
At the system level, page coloring provides cache isolation in software~\cite{Raj2009,Kim2012}. Zhang~\etal\cite{Zhang2013} proposed a more relaxed isolation like repeated cache cleansing. These solutions cause performance issues, as they prevent optimal use of the cache.
Application-level countermeasures seek to find the source of information leakage and patch it~\cite{Brickell2006}. However, application-level countermeasures are bounded and cannot prevent cache attacks such as covert channels and Rowhammer. 
In contrast with prevention solutions that incur a loss of performance, using performance counters does not prevent attacks but rather detect them without overhead.

	\subsection{Usage of Hardware Performance Counters in Security}

Hardware performance counters are made for performance monitoring, but security researchers found other applications.
In defensive cases, performance counters allow detection of malware~\cite{Demme2013}, integrity checking of programs~\cite{Malone2011}, control flow integrity~\cite{Xia2012}, and binary analysis~\cite{Willems2012}.
In offensive scenarios, it has been used for side-channel attacks against AES~\cite{Uhsadel2008} and RSA~\cite{Bhattacharya2015}. Performance counters have also been used by Maurice~\etal\cite{Maurice2015RAID} to reverse engineer the complex addressing function of the last-level cache of modern Intel CPUs.

	\subsection{Cache Covert Channels}

Cache covert channels are a well-known problem, and have been studied relatively to the recent evolutions in microarchitecture. The two main types of access-driven attacks can be used to derive a covert channel. Covert channels using \PrimeProbe have already been demonstrated in~\cite{Maurice2015C5,Liu2015}. \FlushReload has been used for side-channels attacks~\cite{Yarom2014}, thus a covert channel can be derived easily. However, to the best of our knowledge, there was no study of the performance of such a covert channel.

In addition to building a covert channel with our new attack \FlushOnly, we re-implemented \PrimeProbe and implemented \FlushReload.\footnote{After public disclosure of the \FlushOnly attack on November 14, 2015, \FlushOnly has also been demonstrated on ARM-based mobile devices~\cite{LippGSM15}.} We thus provide an evaluation and a fair comparison between these different covert channels, in the same hardware setup and with the same protocol.

	\subsection{Side-Channel Attacks on User Inputs}
	
Section~\ref{sec:keystrokes} describes a side channel to eavesdrop on keystrokes.
If an attacker has root access to a system, there are simple ways to implement a keylogger.
Without root access, software-based side-channel attacks have already proven to be a reliable way to eavesdrop on user input.
Attacks exploit the execution time~\cite{Tannous2008}, peaks in CPU and cache activity graphs~\cite{Ristenpart2009}, or system services~\cite{Zhang2009}.
Zhang~\etal\cite{Zhang2009} showed that it is possible to derive key sequences from inter-keystroke timings obtained via \texttt{procfs}.
Oren~\etal\cite{Oren2015} demonstrated that cache attacks in sandboxed JavaScript inside a browser can derive user activities, such as mouse movements. Gruss~\etal\cite{Gruss2015Template} showed that auto-generated \FlushReload attacks can be used to measure keystroke timings as well as identifying keys.

\section{Conclusion} \label{sec:conclusion}
In this paper we presented \FlushOnly, a novel cache attack that, unlike any other, performs no memory accesses. Instead, it relies only on the execution time of the flush instruction to determine whether data is cached.
\FlushOnly does not trigger prefetches and thus is applicable in more situations than other attacks.
The \FlushOnly attack is faster than any existing cache attack. It achieves a transmission rate of 496\,KB/s in a covert channel scenario, which is $6.7$ times faster than any previous cache covert channel. As it performs no memory accesses, the attack causes no cache misses at all. For this reason, detection mechanisms based on performance counters to monitor cache activity fail, as their underlying assumption is incorrect.

While the \FlushOnly attack is significantly harder to detect than existing cache attacks, it can be prevented with small hardware modifications. Making the \clflush instruction constant-time has no measurable impact on today's software and does not introduce any interface changes. Thus, it is an effective countermeasure that should be implemented. 

Finally, the experiments led in this paper broaden the understanding of the internals of modern CPU caches. Beyond the adoption of detection mechanisms, the field of cache attacks benefits from these findings, both to discover new attacks and to be able to prevent them.

\section{Acknowledgments}
We would like to thank Mathias Payer, Anders Fogh, and our anonymous reviewers for their valuable comments and suggestions.

\noindent\begin{tabular}{m{\dimexpr 0.13\hsize} m{2pt} m{\dimexpr 0.87\hsize-5\tabcolsep-2pt}}
\vspace*{1mm}\includegraphics[width=\hsize]{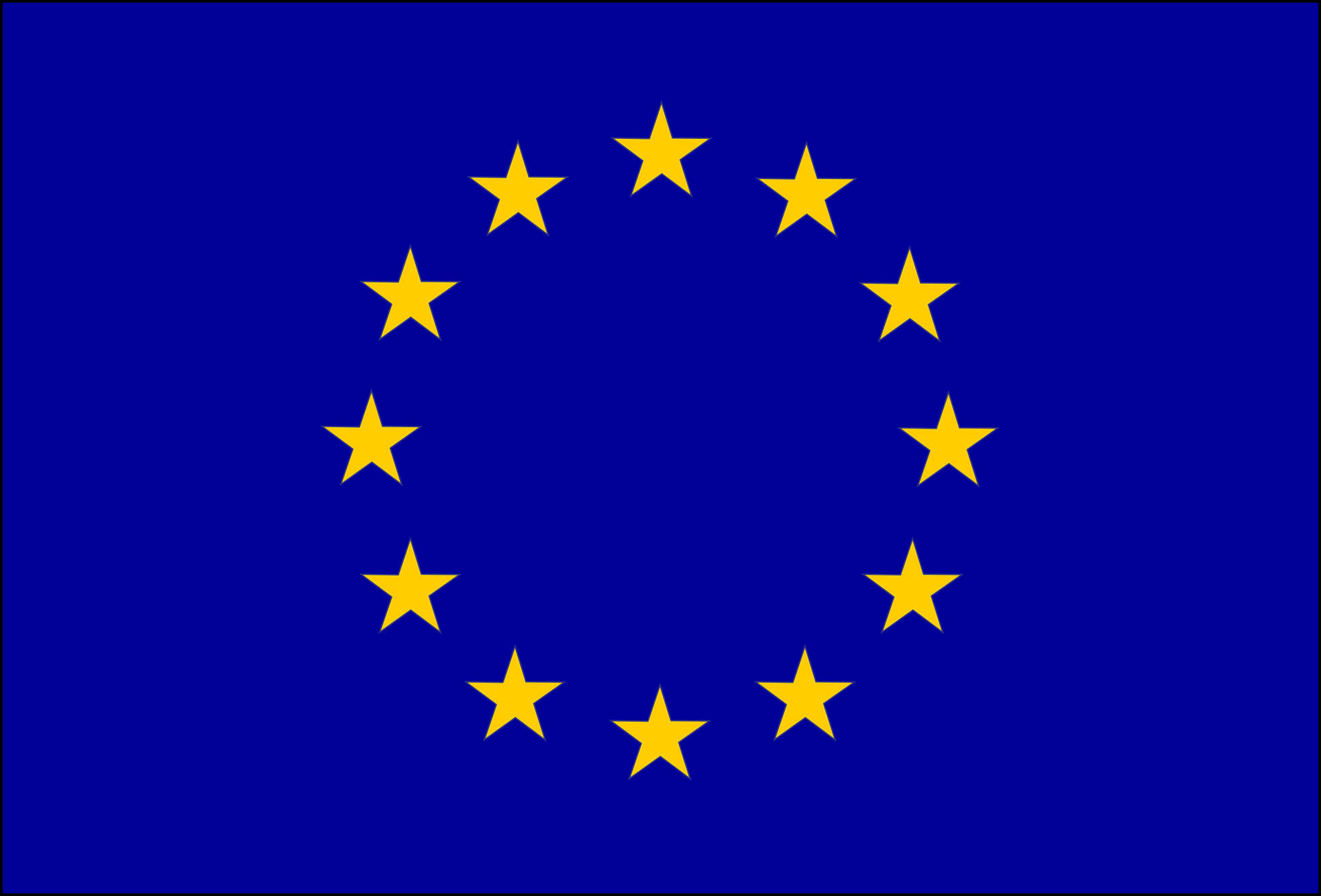}& &Supported by the EU Horizon 2020 programme under GA No. 644052 (HECTOR), the EU FP7 programme under GA No. 610436 (MATTHEW), the Austrian Research Promotion Agency (FFG) \quad \quad \quad \vspace{-\baselineskip}
\end{tabular} \\
and Styrian Business Promotion Agency (SFG) under GA No. 836628 (SeCoS), and Cryptacus COST Action IC1403.

{\footnotesize \bibliographystyle{splncs}
\bibliography{biblio}}

\end{document}